\documentclass[letter,tradiabstract,longauth]{aa}

\usepackage{txfonts}
\usepackage{graphicx}
\usepackage{float}
\usepackage{natbib}
\usepackage[english]{babel}
\usepackage{color}
\usepackage[flushleft]{threeparttable}

\begin{document}
\title{An ALMA view of the interstellar medium of the z=4.77 lensed starburst SPT-S J213242-5802.9}


\def\ESOGarching{1}
\def\Diego{2}
\def\Cambridge{3}
\def\Dal{4}
\def\UFlorida{5}
\def\UCL{6}
\def\UCLA{7}
\def\Arizona{8}
\def\IPAC{9}
\def\CfA{10}
\def\MPIfR{11}
\def\Illinois{12}
\def\Oxford{13}


\author{M.~B\'ethermin$^{\ESOGarching}$\thanks{mbetherm@eso.org},
C.~De~Breuck$^{\ESOGarching}$,
B.~Gullberg$^{\ESOGarching}$,
M.~Aravena$^{\Diego}$,
M.~S.~Bothwell$^{\Cambridge}$,
S.~C.~Chapman$^{\Dal}$,
A.~H.~Gonzalez$^{\UFlorida}$,
T.~R.~Greve$^{\UCL}$,	
K.~Litke$^{\Arizona}$,
J.~Ma$^{\UFlorida}$,
M.~Malkan$^{\UCLA}$,
D.~P.~Marrone$^{\Arizona}$,
E.~J.~Murphy$^{\IPAC}$,
J.~S.~Spilker$^{\Arizona}$,
A.~A.~Stark$^{\CfA}$,
M.~Strandet$^{\MPIfR}$,
J.~D.~Vieira$^{\Illinois}$,
A.~Wei\ss$^{\MPIfR}$,
N.~Welikala$^{\Oxford}$}

\institute{
$^{\ESOGarching}${European Southern Observatory, Karl Schwarzschild Stra\ss e 2, 85748 Garching, Germany}\\
$^{\Diego}${N\'ucleo de Astronom\'{\i}a, Facultad de Ingenier\'{\i}a, Universidad Diego Portales, Av. Ej\'ercito 441, Santiago, Chile}\\
$^{\Cambridge}${Cavendish Laboratory, University of Cambridge, JJ Thompson Ave, Cambridge CB3 0HA, UK}\\
$^{\Dal}${Dalhousie University, Halifax, Nova Scotia, Canada}\\
$^{\UFlorida}${Department of Astronomy, University of Florida, Gainesville, FL 32611, USA}\\
$^{\UCL}${Department of Physics and Astronomy, University College London, Gower Street, London WC1E 6BT, UK}\\
$^{\UCLA}${Department of Physics and Astronomy, University of California, Los Angeles, CA 90095-1547, USA}\\
$^{\Arizona}${Steward Observatory, University of Arizona, 933 North Cherry Avenue, Tucson, AZ 85721, USA}\\
$^{\IPAC}${Infrared Processing and Analysis Center, California Institute of Technology, MC 220-6, Pasadena, CA 91125, USA}\\
$^{\CfA}${Harvard-Smithsonian Center for Astrophysics, 60 Garden Street, Cambridge, MA 02138, USA}\\
$^{\MPIfR}${Max-Planck-Institut f\"{u}r Radioastronomie, Auf dem H\"{u}gel 69 D-53121 Bonn, Germany}\\
$^{\Illinois}${Department of Astronomy and Department of Physics, University of Illinois, 1002 West Green Street, Urbana, IL 61801, USA}\\
$^{\Oxford}${Department of Physiscs, Oxford University, Denis Wilkinson Building, Keble Road, Oxford, OX1 3RH, UK}\\
}

\date{Received ??? / Accepted ???}

\abstract{We present ALMA detections of the [NII] 205\,$\mu$m and CO(12-11) emission lines, and the tentative detection of [CI] $^3$P$_1$ - $^3$P$_0$ for the strongly lensed ($\mu$=5.7$\pm$0.5) dusty, star-forming galaxy SPT-S J213242-5802.9 (hereafter SPT2132-58) at z=4.77. The [NII] and CO(12-11) lines are detected at 11.5 and 8.5\,$\sigma$ level, respectively, by our band-6 observations. The [CI] line is detected at 3.2\,$\sigma$ after a re-analysis of existing band-3 data. The [CI] luminosity implies a gas mass of 3.8$\pm$1.2$\times$10$^{10}$\,M$_\odot$, and consequently a very short depletion timescale of 34$\pm$13\,Myr and a CO-luminosity-to-gas-mass conversion factor $\alpha_{\rm CO}$ of 1.0$\pm$0.3\,M$_\odot$ (K km s$^{-1}$ pc$^{2}$)$^{-1}$. SPT2132-58 is an extreme starburst with an intrinsic star formation rate of 1100$\pm$200\,M$_\odot$/yr. We find a [CII]/[NII] ratio of 26$\pm$6, which is the highest reported at z$>$4. This suggests that SPT2132-58 hosts an evolved interstellar medium ($\rm 0.5\,Z_\odot<Z<1.5\,Z_\odot$), which may be dominated by photodissociation regions. The CO(2-1) and CO(5-4) transitions have lower CO-to-far-infrared ratios than local and high-redshift samples, while CO(12-11) is similar to these samples, suggesting the presence of an additional very excited component or an AGN.} 

\keywords{Galaxies: starburst -- Galaxies: ISM --  Galaxies: high-redshift -- Galaxies: star formation -- Submillimeter: galaxies}

\titlerunning{An ALMA view on the ISM of the z=4.77 lensed starburst SPT2132-58}

\authorrunning{B\'ethermin et al.}

\maketitle

\section{Introduction}

The presence of a large population of massive, dusty galaxies hosting high star formation rates (SFR$>$500\,M$_\odot$/yr) at high redshift \citep[e.g.,][]{Smail1997,Blain2002,Chapman2005} cannot be explained by the standard semi-analytical models of galaxy formation and hydrodynamical simulations \citep[e.g.,][]{Cousin2015a,Sparre2015}. Recently, several of these extreme star-forming galaxies have been found at z$>$4 \citep[e.g.,][]{Riechers2013,Vieira2013,Dowell2014,Smolcic2015}, indicating extremely quick assembly of these structures. Their high SFRs suggest the presence of large gas reservoirs and their large dust content indicates that their interstellar medium (ISM) is metal enriched. This raises the question of how can such a massive and mature ISM be built so early?

Studying the ISM of these z$>$4 galaxies is difficult using the rest-frame optical, because of the strong dust attenuation \citep{DaCunha2015}. The dust continuum and the (sub-)millimeter lines from the ISM are thus the main information we can obtain and the continuum and [CII] can now be routinely detected by ALMA \citep[e.g.,][]{Scoville2014,Capak2015}. The [CII]/[NII] line ratio was proposed as a new diagnostic of the ISM in star-forming galaxies \citep{Nagao2012}, but its use is hampered by the faintness of the [NII] line requiring long integration times, even with ALMA \citep{Nagao2012,Decarli2014}. This problem can be solved by targeting sources amplified by gravitational lensing. The sample of strongly-lensed sources found by the South Pole Telescope \citep[SPT, ][]{Carlstrom2011} is particularly suited for these follow-up studies because of its high median redshift ($\langle$z$\rangle$=3.9, \citealt{Strandet2015}).

In this Letter, we report the detections of the [NII], CO(12-11), and [CI] $^3$P$_1$ - $^3$P$_0$ lines in SPT-S J213242-5802.9 at $z$=4.768 \citep[hereafter, SPT2132-58;][]{Weiss2013,Gullberg2015}. We assume the \citet{Planck_cosmo2015} cosmology (H$_0$ = 67.8\,km s$^{-1}$ Mpc$^{-1}$, $\Omega_{\rm m}$ = 0.308).

\begin{figure*}
\centering
\begin{tabular}{cc}
\includegraphics[width=8cm]{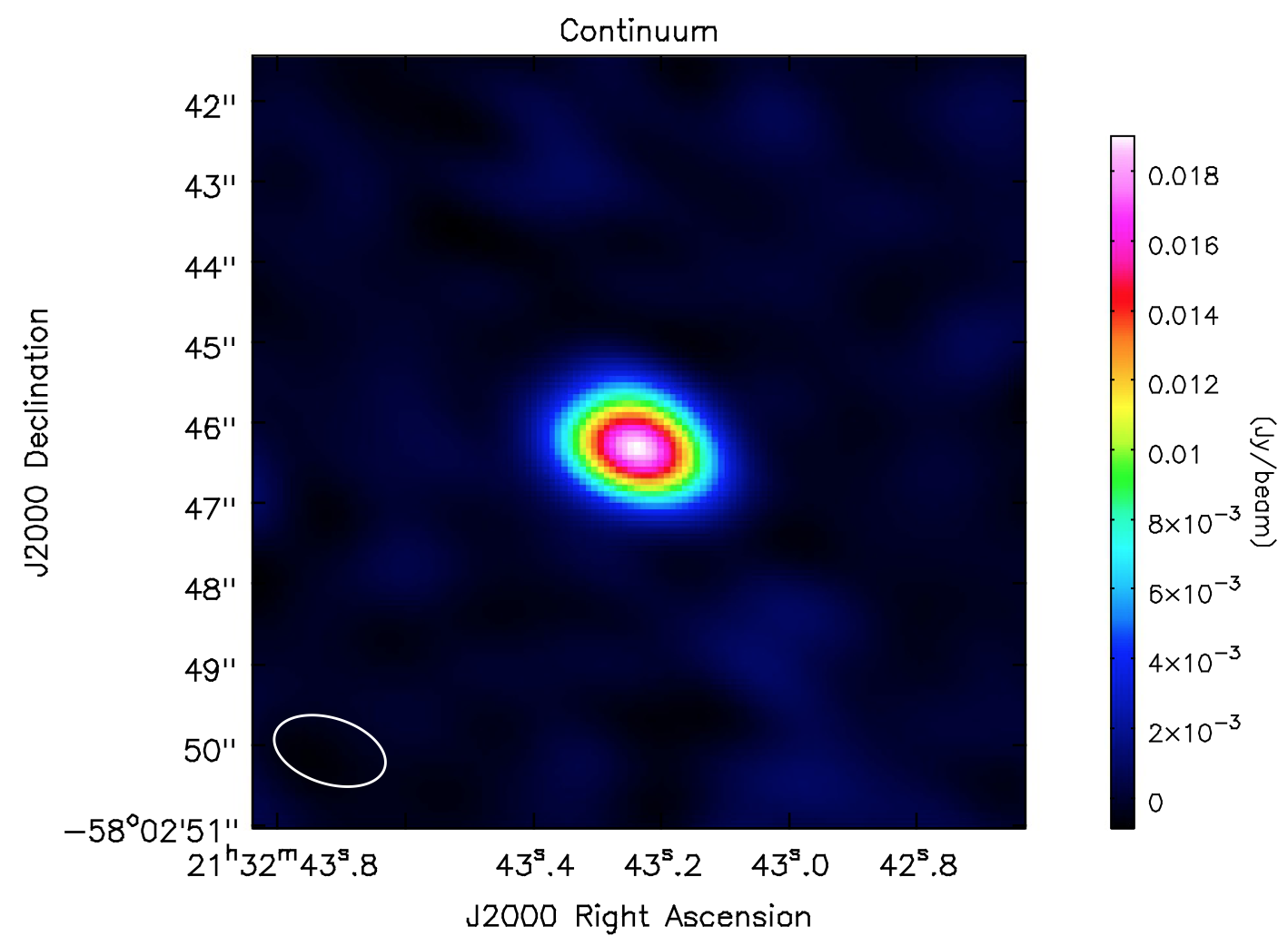} & \includegraphics[width=8cm]{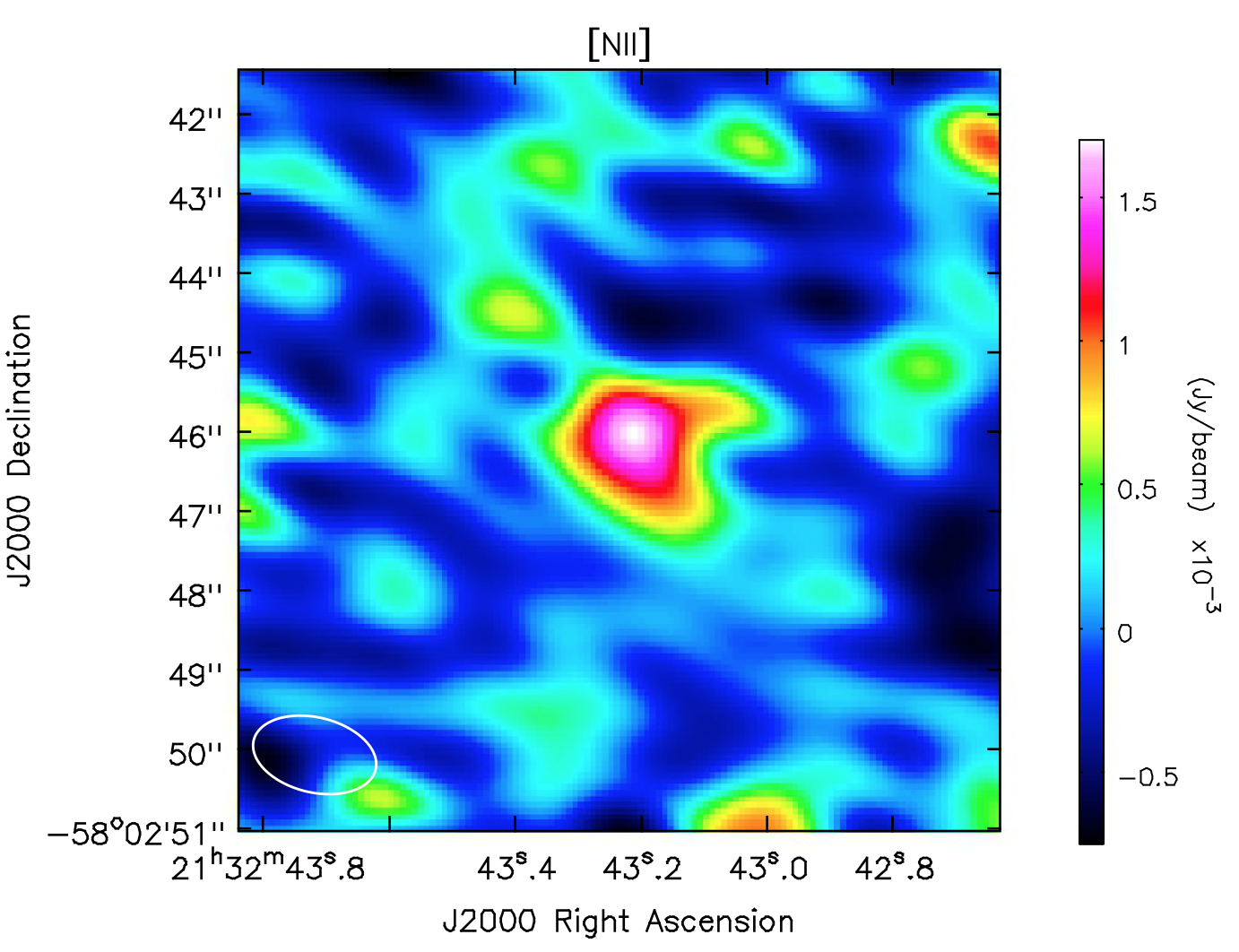}\\
\includegraphics[width=8cm]{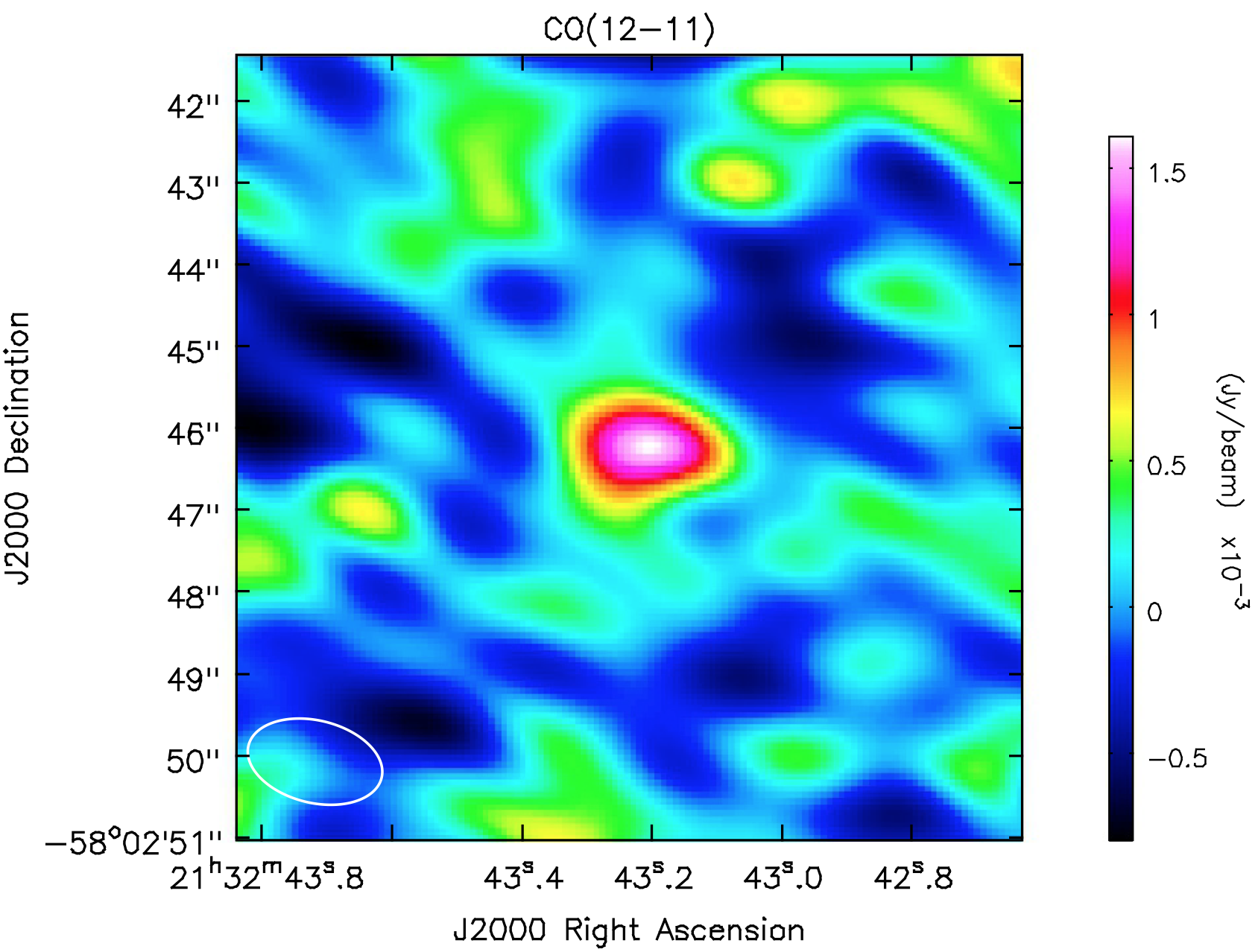} & \includegraphics[width=8cm]{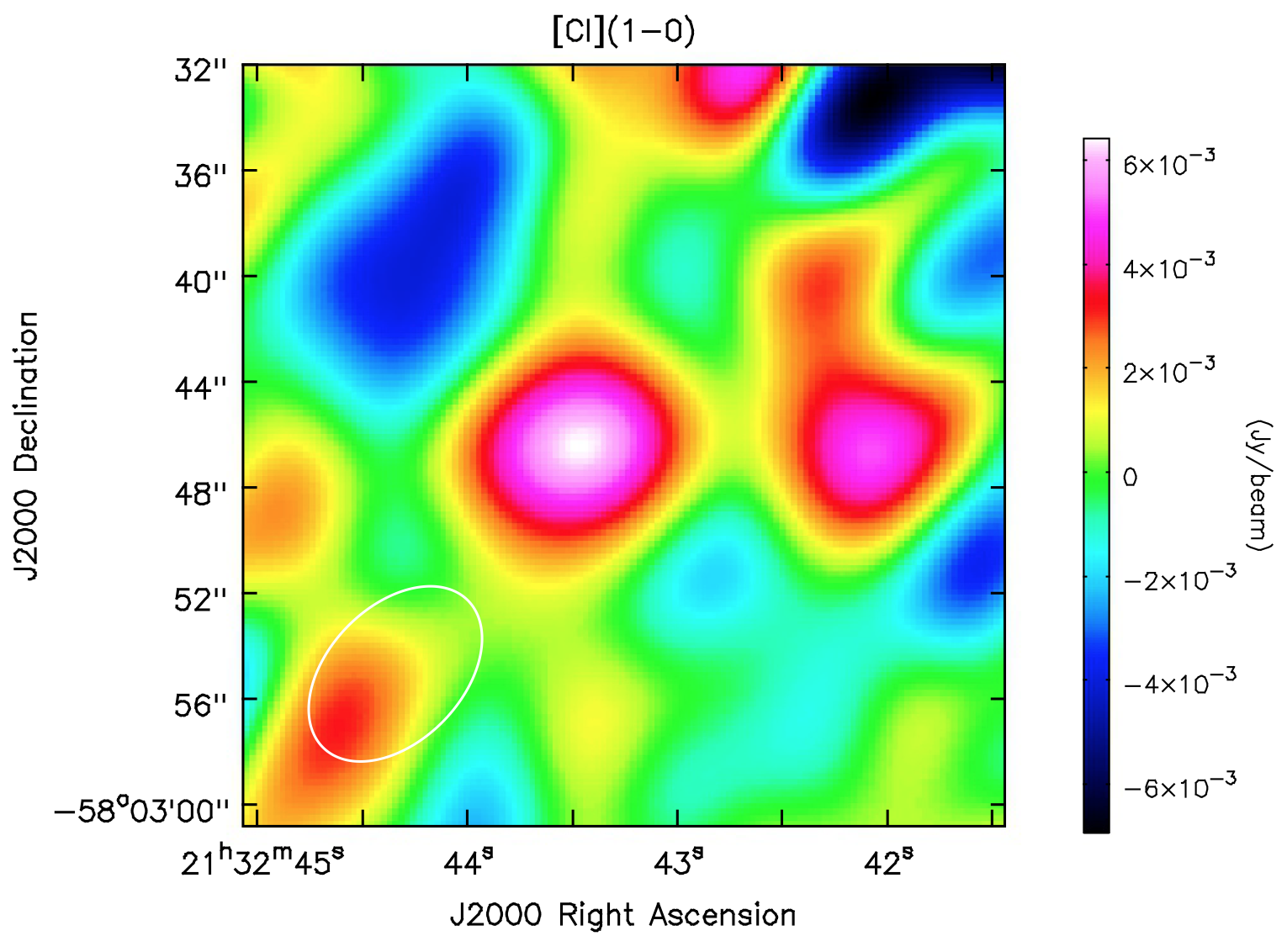} \\
\end{tabular}
\caption{\label{fig:maps} Maps of the continuum (top left), the [NII] line (top right), the CO(12-11) line (bottom left) and the [CI](1-0) line (bottom right). The size of the beam is indicated in the lower left corner. [NII] and CO(12-11) were integrated between -300 and +300 km/s. Since the [CI](1-0) line is narrower and fainter, it was integrated between -100 and +100 km/s to maximize the signal-to-noise ratio. The beam in band 3 ([CI](1-0)) is much larger than in band 6. We thus stretch the x and y axis by a factor 3.}
\end{figure*}

\section{Data}

\label{sect:data}

SPT2132-58 was observed with ALMA in band 6 on February 23 2014 (2012.1.00994.S, PI: De Breuck) with a relatively compact configuration (minimal baseline = 15\,m, max = 399\,m, RMS = 140\,m) and good weather (PWV = 0.9\,mm). The on-source integration time is 470\,s. The data were calibrated by the standard ALMA pipeline. The four spectral windows are centered on 237.6, 239.5, 252.9, and 254.8\,GHz. Each spectral window has 480 channels and covers 1875\,MHz. We produced the images of our source with CASA \citep{McMullin2007} using the \texttt{clean} routine, and we chose a natural weighting to optimize the sensitivity (see Sect.\,\ref{sect:results}). This results in a synthesized beam size of 1.55"$\times$0.91". The typical RMS noise is 0.36\,mJy/beam for the continuum (combining the 4 spectral windows) and 1.6\,mJy/beam/(25 km/s channel) for the spectra. The spectra were extracted using an elongated aperture 2 times larger in diameter than the beam to avoid losing flux because our source is marginally resolved. We also produced a continuum image using a Briggs weighting with robust = 0.5, since the signal-to-noise ratio is already very high (SNR = 52, see Fig.\,\ref{fig:maps}).

In addition, we reanalyzed band 3 data obtained during the cycle 0 \citep{Weiss2013} using the same approach (see Fig.\,\ref{fig:maps}). We used a natural weighting instead of a Briggs weighting with robust=0.5. The beam size is 7.7"$\times$4.9" and the noise is 3.2\,mJy/beam/(50 km/s channel). The natural weighting reduces the noise in the spectrum and improves the signal-to-noise ratio of the [CI] line from 2.9 to 3.2.

\begin{table*}
\caption{\label{tab:summary}Summary of the properties of the detected lines of SPT2132-58. These apparent luminosities have not been corrected for the lensing magnification.}
\centering
\begin{tabular}{lllllcll}
\hline
\hline
Line & $\nu_{\rm rest}$ & $I_{\rm line}$ \tablefootmark{a} & $\mu$\,L'$_{\rm line}$ & $\mu$\,L$_{\rm line}$ & FWHM & Instrument & Reference \\
 & GHz & Jy\,km/s & 10$^{10}$\,K km$^{-1}$ pc$^2$ & 10$^8$\,L$_\odot$ & km/s & & \\
\hline
CO(2-1) & 230.54 & 0.85$\pm$0.07 & 18.38$\pm$1.51 & 0.71$\pm$0.06 & 225$\pm$17 & ATCA & \citet{Aravena2015}\\
$[$CI(1-0)$]$ & 492.16 & 0.86$\pm$0.27 & 4.07$\pm$1.29 & 1.54$\pm$0.49 & ... & ALMA band 3 & This work\\
CO(5-4) & 576.35 & 1.41$\pm$0.26 & 4.88$\pm$0.90 & 2.96$\pm$0.54 & 231$\pm$70 & ALMA band 3 & \citet{Weiss2013}\\
CO(12-11) & 1383.24 & 1.36$\pm$0.16 & 0.82$\pm$0.10 & 6.84$\pm$0.81 & 331$\pm$78 & ALMA band 6 & This work\\
$[$NII$]$ & 1461.10 & 1.73$\pm$0.15 & 0.93$\pm$0.08 & 9.20$\pm$0.79 & 240$\pm$41 & ALMA band 6 & This work\\
$[$CII$]$ & 1900.54 & 34.90$\pm$6.90 & 11.10$\pm$2.19 & 241.53$\pm$47.75 & 212$\pm$43 & APEX/FLASH & \citet{Gullberg2015}\\
\hline
\end{tabular}
\tablefoot{
\tablefoottext{a}{The integrated line flux $I_{\rm CO}$ is computed between -1.5 and 1.5 FWHM.}
}
\end{table*}

\begin{figure}
\centering
\includegraphics[width=9cm]{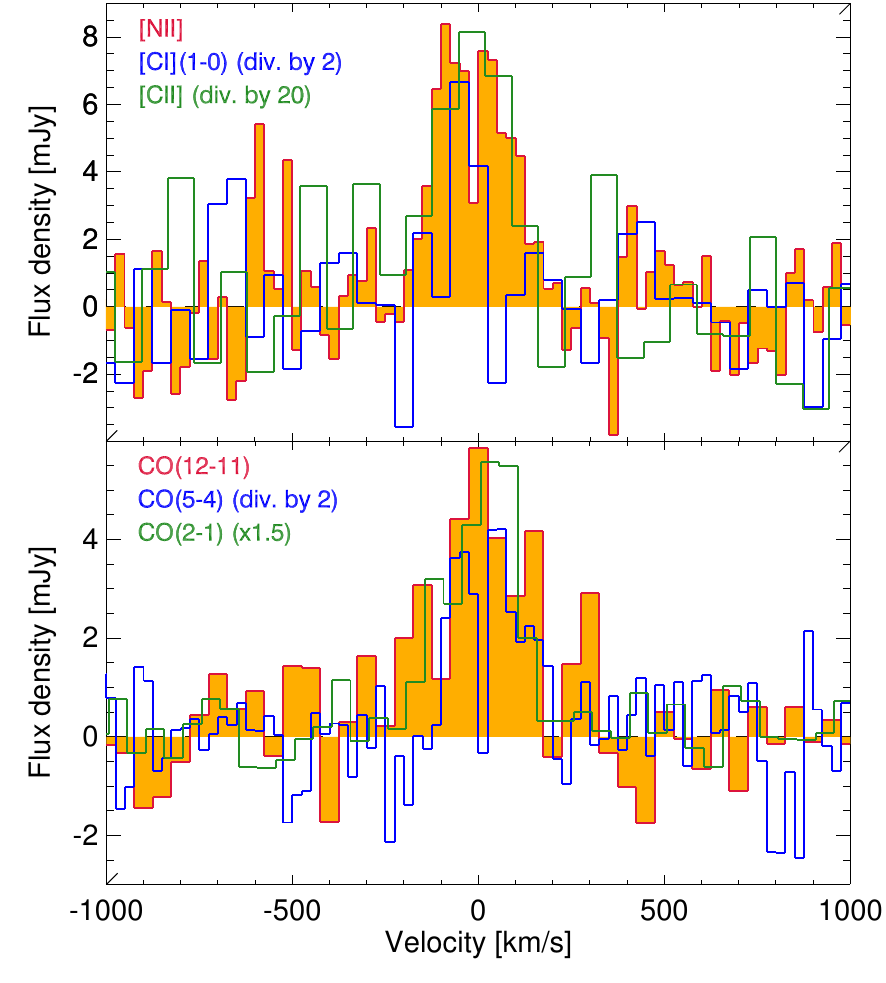}
\caption{\label{fig:profiles} \textit{Upper panel:} Profile of the fine structure lines of SPT2132-58 ([NII] in orange, [CI](1-0) in blue and [CII] in green). \textit{Lower panel:} Profile of the CO lines (12-11 in orange, 5-4 in blue and 2-1 in green).}
\end{figure}

\section{Results}

\label{sect:results}

The [NII] and CO(12-11) lines are detected unambiguously at 11.5 and 8.5\,$\sigma$, respectively (see Fig.\,\ref{fig:profiles}). The re-analysis of the cycle 0 data allowed us to obtain a tentative detection of [CI] at 3.2\,$\sigma$. The fluxes of the lines are listed in Table\,\ref{tab:summary}. Finally, we measured the flux density of the source and found an integrated flux 26.4$\pm$0.5\,mJy at 246\,GHz. The [NII] and CO(12-11) lines were not excluded from this measurement, since ancillary bolometric data also include lines and their contribution is only 1.3\% of the total continuum flux integrated over the four spectral windows. The source is marginally resolved along its major axis. AIPS and CASA provide a similar best estimate of the source size of (0.6$\pm$0.2)". This value is consistent with the size in the image plane predicted by our best lens model (0.67", Spilker et al. in prep.).

We compared the profiles of these new detections with the previously reported detections of CO(5-4) by \citet{Weiss2013}, [CII] by \citet{Gullberg2015}, and CO(2-1) by \citet{Aravena2015}. All the lines are compatible with a single Gaussian function (highest reduced $\chi^2$ is 1.4, which is compatible at 1\,$\sigma$ with unity considering their number of degrees of freedom). The results of the fits are given in Table\,\ref{tab:summary}. The [NII], [CII], CO(5-4) and CO(2-1) lines have similar full widths at half maximum (FWHMs). The CO(12-11) line is slightly wider than the other lines (1.0\,$\sigma$ compared to [NII]). The low SNR of [CI] prohibits us from determining the FWHM.

\begin{figure}
\centering
\includegraphics[width=9cm]{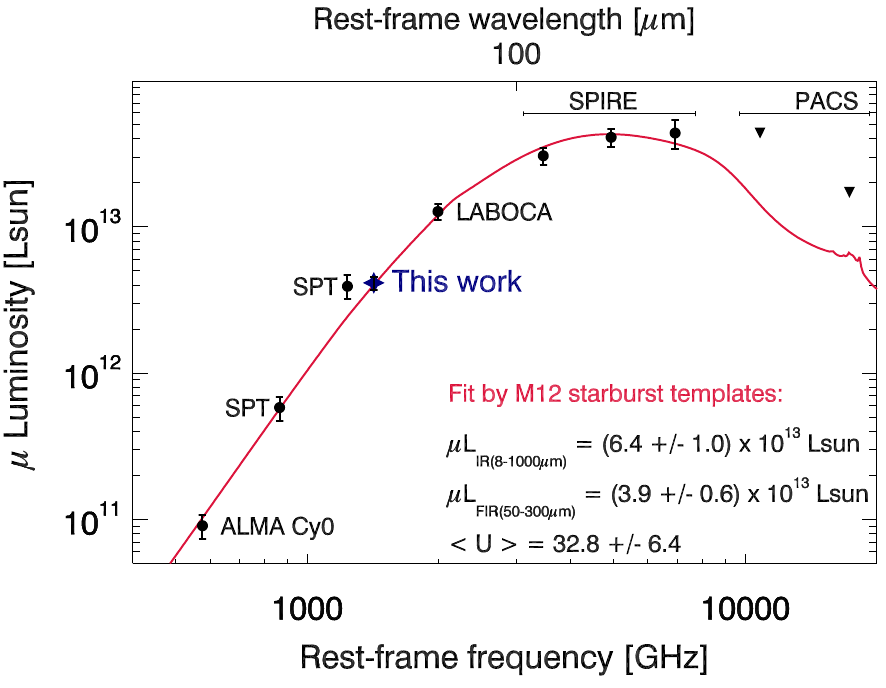}
\caption{\label{fig:sed} SED of SPT2132-58, including a compilation of measurements from Strandet et al. in prep. (filled black circles for detections, downward triangles for 3$\sigma$ upper limits) and our new ALMA Cycle 1 continuum measurement (blue filled star), and best fit by the starburst templates of the \citet{Magdis2012b} model (red line). The $\langle$U$\rangle$ parameter is the mean intensity of the radiation field heating the dust compared to the solar neighborhood.}
\end{figure}

\section{Discussion}

\label{sect:discussion}

\subsection{Star formation efficiency and nature of the source}

\label{sect:sfe}

SPT2132-58 is gravitationally magnified by another galaxy on the line of sight. Our best-model based on the ALMA cycle-0 high-resolution follow-up of the source estimates the magnification to be $\mu$=5.7$\pm$0.5 (Spilker et al. in prep.). We determined its apparent total infrared luminosity ($\mu$\,L$_{\rm IR(8-1000\,\mu m)}$ = (6.4$\pm$1.0)$\times$10$^{13}$\,L$_\odot$) and the far-infrared luminosity ($\mu$\,L$_{\rm FIR(50-300\,\mu m)}$ = (3.9$\pm$0.6)$\times$10$^{13}$\,L$_\odot$) by fitting the continuum measurements made by ALMA band 3, SPT, LABOCA, \textit{Herschel}/SPIRE (Strandet et al. in prep.), and ALMA band 6 (Sect.\,\ref{sect:results} and Fig.\,\ref{fig:sed}). We used the starburst templates of \citet{Magdis2012b} to fit the continuum. The quality of the fit is good ($\chi^2_{\rm red}$=1.1), but the SPT point at 220\,GHz exhibits a 2\,$\sigma$ excess. This tension could be explained by a peak of noise at the position of our source and not an astrophysical contaminant since no excess is found in the other SPT band. We found an apparent luminosity $\mu$\,L$_{\rm IR}$ of (6.4$\pm$1.0)$\times$10$^{13}$\,L$_\odot$. Because of the high dust temperature of the source (T$_{\rm dust}$=41K, \citealt{Aravena2015}), the CMB has a negligible impact on this measurement \citep{Da_Cunha2013}. The intrinsic L$_{\rm IR}$ is thus around 10$^{13}$\,L$_\odot$ putting this source at the edge of the hyper luminous infrared galaxy (HyLIRG, L$_{\rm IR}>10^{13}$\,L$_\odot$) category. This corresponds to an intrinsic SFR of 1120$\pm$200\,M$_\odot$/yr (\citealt{Kennicutt1998} assuming a \citealt{Chabrier2003} initial mass function). This SFR range is expected to be only populated by episodic starbursts according to the \citet{Bethermin2012c} model.

We estimated the gas mass from the [CI] emission \citep{Papadopoulos2004}. Assuming the abundance and excitation factors of \citet{Alaghband_Zadeh2013}, we found an apparent gas mass $\mu$\,M$_{\rm gas}$ of (21.5$\pm$6.7$)\times$10$^{10}$\,M$_\odot$ and a demagnified gas mass of (3.8$\pm$1.2)$\times$10$^{10}$\,M$_\odot$. A systematic analysis of all [CI] lines detected in SPT sources and a detailed discussion about the accuracy of [CI]-derived gas mass will be presented in Bothwell et al. (in prep.). We can also estimate the $\alpha_{\rm CO}$ of this object from the L$'_{\rm CO(2-1)}$ line luminosity measured by \citet{Aravena2015}, converting it into L$'_{\rm CO(1-0)}$ using the mean SMG ratio \citep{Carilli2013}, and assuming the gas mass provided by [CI]. We find $\alpha_{\rm CO}$ = 0.99$\pm$0.32\,M$_\odot$\,(K km s$^{-1}$ pc$^{2}$)$^{-1}$. This agrees with $\alpha_{\rm CO}$ of 0.9$\pm$0.3\,(K km s$^{-1}$ pc$^{2}$)$^{-1}$ derived by \citet{Aravena2015} from the dust mass assuming a gas-to-dust mass ratio of 100. This value is compatible with the classical $\sim$0.8\,M$_\odot$\,\,(K km s$^{-1}$ pc$^{2}$)$^{-1}$ found for local compact starbursts \citep{Downes1998}.

The depletion timescale associated with this gas mass and SFR is 34$\pm$13\,Myr. This is a factor of 3 shorter than typical strong starbursts and more than an order of magnitude less than normal star-forming galaxies at this redshift \citep[e.g.,][]{Bethermin2015}. The short depletion timescale indicates that this source is an extreme starburst with a moderate gas content, but a very high star formation efficiency. The depletion timescale could however be slightly underestimated if this source is observed at the end of the starburst. In this case, the SFR would be quickly decreasing. The SFR derived from L$_{\rm IR}$ would thus be lower than the actual SFR, since this tracer is sensitive to the SFR in the last 10-100\,Myr \citep{Kennicutt1998}.

\begin{figure*}
\centering
\includegraphics[width=15cm]{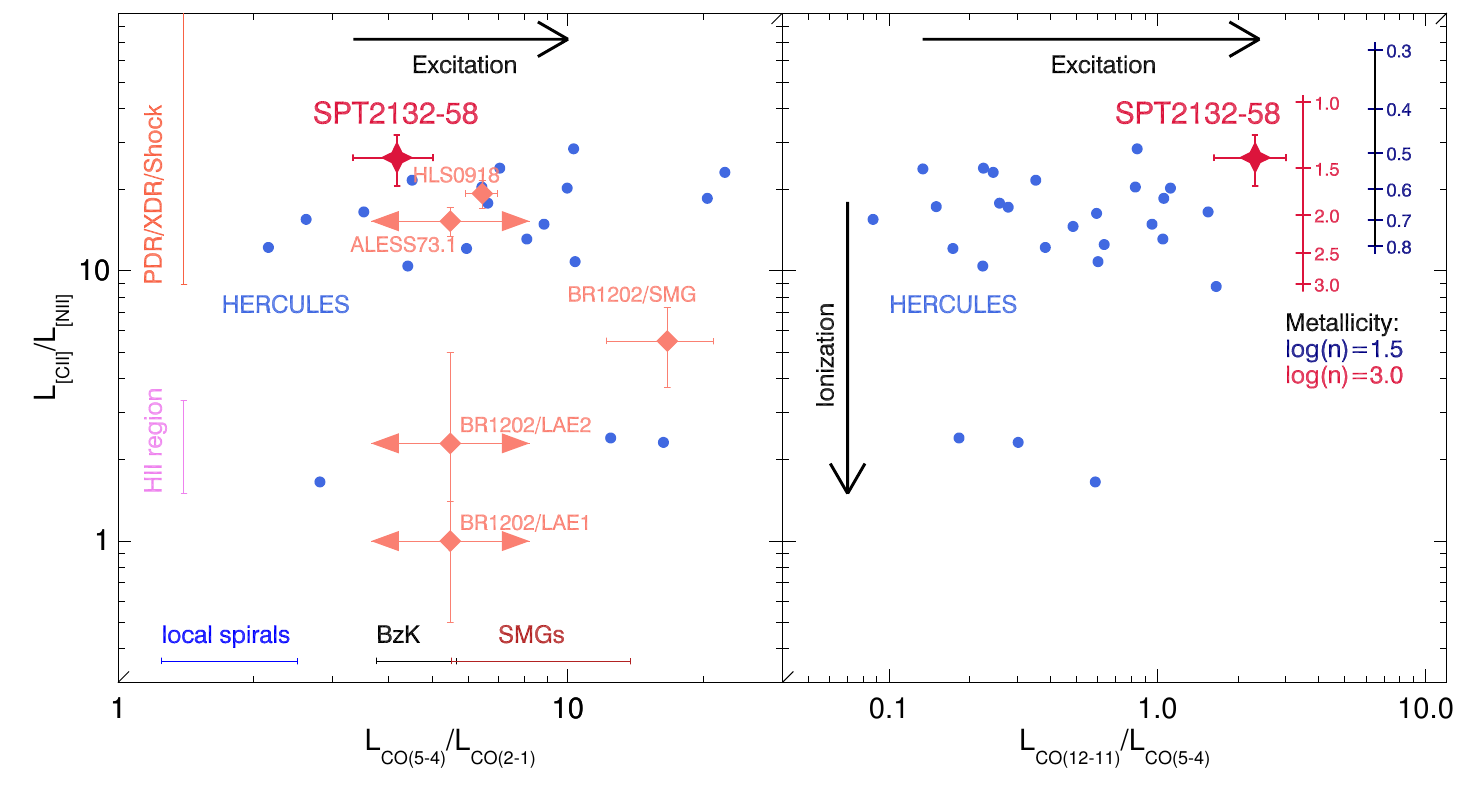}
\caption{\label{fig:line_ratios} CO excitation (CO(5-4)/CO(2-1) in left panel and CO(12-11)/CO(5-4) in the right panel) versus [CII]/[NII] luminosity ratio (the ratio reported for ALESS73.1 is computed using the updated [CII] flux of \citealt{DeBreuck2014}). Red star: SPT2132-58. Salmon diamonds: other high-z sources from the literature (see text). Blue circles: HERCULES local sample \citep{Rosenberg2015}. [NII] luminosities are from \citet{Kamenetzky2015}. The left and right arrows indicate that this source does not have available measurement of its CO ratio. The range of CO(5-4)/CO(2-1) ratio for various sample is indicated in the bottom of the left panel \citep{Daddi2015}. The ranges of ratio expected for HII and PDR/XDR/Shock dominated regions from \citet{Decarli2014} are indicated on the left. The metallicity estimated from [CII]/[NII] using the \citet{Nagao2012} method is indicated on the right for log(n)=1.5 (navy) and log(n)=3.0 (blue). We assume log(U$_{\rm HII}$)=-3.5 for both.}
\end{figure*}

\begin{figure}
\centering
\includegraphics[width=9cm]{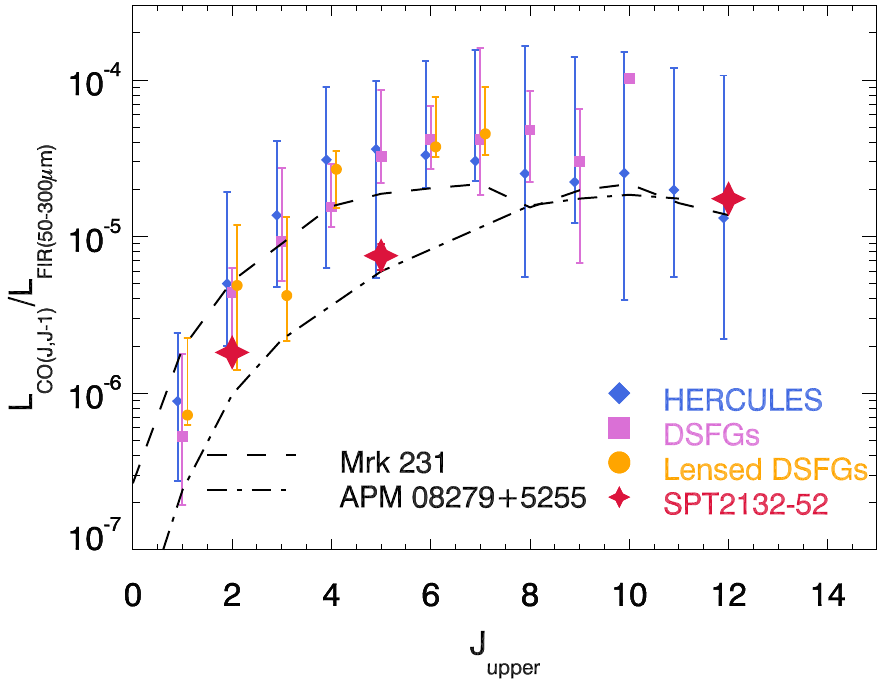}
\caption{\label{fig:sled} CO spectral line energy distribution (SLED) of SPT2132-58 normalized by L$_{\rm FIR}$. Our results are compared with three data compilations of \citet{Greve2014} (the error bars indicate the lowest and the highest value of their samples). The dashed line is the SLED of Mrk231 (\citealt{Rosenberg2015}) and the dot-dash line is the SLED of APM 08279 +5255 (\citealt{Downes1999,Weiss2007,Riechers2009,Bradford2011}).}
\end{figure}

\subsection{ISM properties}

SPT2132-58 is one of the few high-redshift sources that have been detected in both [CII] and [NII] 205\,$\mu$m (y-axis of Fig.\,\ref{fig:line_ratios}). The previous detections were obtained in HLS~J091828.6+514223 \citep{Combes2012,Rawle2014}, ALESS73.1 \citep{Nagao2012}, and one SMG and two LAEs (tentative detections) in the BR1202-0725 system \citep{Decarli2014}. We find a [CII]/[NII] luminosity ratio of 26.3$\pm$5.7 in SPT2132-58, which is the highest reported so far at high-redshift. HLS~J0918 has a similar ratio within 1\,$\sigma$, but ALESS73.1 is 2 times lower. The members of the BR1202-0725 system have a much lower ratio (4--26 times). Finally, we compared SPT2132-58 with the HERCULES sample of local (U)LIRGs \citep{Rosenberg2015}. All but 3 of the 29  HERCULES sources have a lower ratio. However, the [CII]/[NII] ratio of SPT2132-58 is not uncommon in the local Universe and is close to, e.g., the mean value of 30 found by \citet{Spinoglio2015} in local Seyfert galaxies.

The [CII]/[NII] ratio of SPT2132-58 corresponds to a PDR/XDR/shock-dominated regime according to \citet{Decarli2014}. For galaxies in this regime, the [CII] originates from both HII regions and external layers of PDRs, while [NII] comes only from HII regions. Using a model that takes both PDR and HII regions into account, \citet{Nagao2012} showed that the observed [CII]/[NII] ratio depends on metallicity. This occurs since the PDR/HII fraction of [CII] varies with metallicity. Other parameters affecting this ratio include density and, to a lesser extent, the radiation field. Fig.\,\ref{fig:line_ratios} shows the predictions of \citet{Nagao2012} for both a low (navy) and a high (red) density, but only for a single ionization parameter log(U$_{\rm HII}$)=-3.5. The [CII]/[NII] ratio of SPT2132-58 corresponds to a metallicity in the 0.5 to 1.5\,Z$_\odot$ range. While we could have expected a very low metallicity at such a high redshift, our source is not far from solar metallicity. Nevertheless, our source has a slightly higher [CII]/[NII] ratio than the HERCULES sample. Although the [CII]/[NII] ratio depends on several other physical parameters (e.g., density), the higher ratio found in SPT2132-58 suggests a slightly lower metallicity than in local starbursts.

We also compared the CO excitation with the samples cited previously (x-axis of Fig.\,\ref{fig:line_ratios}). SPT2132-58 has a CO(5-4)/CO(2-1) ratio close to the average of the HERCULES sample. Assuming T$_{\rm kin}$=40\,K, the CMB at z=4.77 has only an impact of 15\% on this ratio \citep{Da_Cunha2013}. The CO(5-4)/CO(2-1) ratio lies at the border between the value found for BzK galaxies and SMGs \citep{Daddi2015} and is thus standard for a high-redshift galaxy. However, these two lines have a lower L$_{\rm CO}$/L$_{\rm FIR}$(50--300\,$\mu$m) ratio than the other samples of DSFGs by a factor of 2 and 3 for CO(2-1) and CO(5-4) transition, respectively (see Fig.\,\ref{fig:sled}). This is compatible with the moderate gas content and the high SFR found in Sect.\,\ref{sect:sfe}, since L$_{\rm CO}$ correlates with the gas content and SFR with L$_{\rm FIR(50-300\,\mu m)}$. The CO(5-4) line is also a factor of 3.2$\pm$0.4 lower than the average CO(5-4) flux of the SPT sample measured by stacking in \citet{Spilker2014}. 

In contrast, the L'$_{\rm CO(12-11)}$/L$_{\rm FIR(50-300\,\mu m)}$ ratio is similar with the HERCULES sample, but the CO(12-11)/CO(5-4) ratio of SPT2132-58 is higher than all the objects of the HERCULES sample. There is thus a deficit of flux of the low and mid-J CO lines, but a strong CO(12-11) line, which may be emitted by different regions and/or excited by another mechanism. This could involve an active galactic nucleus (AGN) and/or regions with a particularly high radiation field. This effect could possibly be amplified by differential magnification \citep{Serjeant2012}. If a very excited compact region is located on a caustic and has its flux magnified by a much larger factor than the rest of the system, the apparent ratio between a high-J and a low-J CO lines can be higher than the  intrinsic ratio. In Fig.\,\ref{fig:sled}, we compare the SLED of SPT2132-58 with the AGN-host galaxies Mrk231 and APM 08279+5255. The SPT2132-58 has a similar L$_{\rm CO(12-11)}$/L$_{\rm FIR}$(50--300\,$\mu$m) ratio to these two objects. At J$\leq$5, there is a factor of 5 of difference between Mrk231, which matches the properties of the HERCULES sample, and APM 08279+5255, which is a strong outlier. SPT2132-58 has intermediate properties and its SLED does not allow us to conclude on the presence or not of an AGN. Future [OI] or mid-infrared observations are required to test the hypothesis.

\section{Conclusion}

We investigated the physical properties of SPT2132-58, a lensed source at z=4.77. We determined that this source is an HyLIRG (SFR=1120$\pm$200\,M$_\odot$/yr) with a relatively modest gas reservoir (3.8$\pm$1.2$\times$10$^{10}$\,M$_\odot$) and a very short gas depletion timescale (34$\pm$13\,Myr), indicating a very intense starburst. The $\alpha_{\rm CO}$ conversion factor (0.99$\pm$0.32) is similar with the one measured in starbursts at lower redshift. The [CII]/[NII] ratio (26.3$\pm$5.7) is the highest found in a high-redshift galaxy and indicates a chemically-evolved ISM (0.5\,Z$_\odot$$<$Z$<$1.5\,Z$_\odot$). This source is an important laboratory to understand the extremely quick stellar mass assembly of the first massive galaxies.

\begin{acknowledgements}
We thanks Roberto Decarli for his very useful suggestions, especially about the choices of axes for Fig\,\ref{fig:line_ratios}, and the anonymous referee for her/his very constructive comments. This paper makes use of the following ALMA data: ADS/JAO.ALMA\#2011.0.00957.S and ADS/JAO.ALMA\#2012.1.00994.S. ALMA is a partnership of ESO (representing its member states), NSF (USA) and NINS (Japan), together with NRC (Canada), NSC and ASIAA (Taiwan), and KASI (Republic of Korea), in cooperation with the Republic of Chile. The Joint ALMA Observatory is operated by ESO, AUI/NRAO and NAOJ. The National Radio Astronomy Observatory is a facility of the National Science Foundation operated under cooperative agreement by Associated Universities, Inc. The SPT is supported by the National Science Foundation through grant PLR- 1248097, with partial support through PHY-1125897, the Kavli Foundation and the Gordon and Betty Moore Foundation grant GBMF 947. JDV, KCL, DPM, and JSS acknowledge support from the U.S. National Science Foundation under grant No. AST-1312950.
\end{acknowledgements}

\bibliographystyle{aa}

\bibliography{biblio}

\end{document}